\documentclass[epjST]{svjour}
\usepackage{graphics}
\usepackage{epsf,epsfig}
\def\xb{\overline{x}}

\def\als{\alpha_s}
\def\vk{{\bf k}_{\perp}}
\def\vbs{{\bf b}}

\begin{document}
\title{Polarized GPDs  in pions and kaons electroproduction. Tranversity effects.}
\author{S.V.Goloskokov\fnmsep\thanks{\email{goloskkv@theor.jinr.ru}} }
\institute{Bogoliubov Laboratory of Theoretical  Physics,
  Joint Institute for Nuclear Research, Dubna, Russia}
\abstract{We analyze the electroproduction of pseudoscalar mesons
within the handbag approach. To investigate these reactions, we
consider  the leading-twist contribution together with the
transversity twist-3 effects that are crucial in the description
of experimental data. Our results on the cross section are in
agreement with experiment. We present our predictions for spin
observables.
} 
\maketitle
\section{Introduction}
\label{intro} We investigate the process of pseudoscalar meson
leptoproduction (PML) at large $Q^2$ within the handbag approach,
where the amplitudes factorize into a hard subprocess and soft
part --Generalized Parton Distributions GPDs \cite{fact}. The hard
subprocess amplitudes are calculated by using the modified
perturbative approach  \cite{sterman} that takes into account
quark transverse degrees of freedom as well as gluonic radiation
condensed in a Sudakov factor.

The PML  was  analyzed in \cite{gk09,gk11}. It was shown that the
leading-twist contribution determined by the polarized GPDs is not
sufficient to describe processes of PML. The essential
contributions from the transversity GPDs are needed to be
consistent with experiment. Within the handbag approach, these
twist-3 effects can be modeled by the transversity GPDs $H_T$,
$\bar E_T$, in conjunction with the twist-3 meson wave function.

In this report we study the cross sections of the pion
leptoproduction in the HERMES and CLAS energy range  on the basis
of the model \cite{gk09,gk11}. Our results are in  good agreement
with experiment. We show that the transversity GPDs lead to a
large transverse cross section for most reactions of the
pseudoscalar meson production. Predictions for spin asymmetries in
the pion leptoproduction are presented as well.

At the end, we present the model results  for the cross section of
the $K^+ \Lambda$ leptoproduction,  which is large due to the
transversity contribution \cite{gk11}, and predictions for the
spin asymmetry in this reaction.

\section{Leptoproduction of pseudoscalar mesons}
Hard exclusive PML amplitudes were studied on the basis of the
handbag approach. The typical contributions are shown in Fig.1. In
the left part of the graph we present the meson pole contribution
which appears for the charge meson production. In  Fig. 1 (right)
the example of the handbag diagram is shown. In the leading twist
the last contribution is expressed in terms of the polarized GPDs
$\tilde H$ and $\tilde E$ whose parameterization can be found in
\cite{gk07q}.

\begin{figure}[h!]
\begin{center}
\epsfig{figure=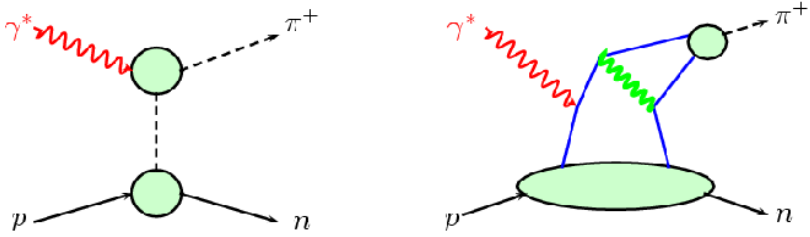, width=10.6cm}
\end{center}
\caption{Examples of the graphs essential in PML. Left--pion pole
and right--handbag  contributions to the  $\pi^+$ production.}
\end{figure}

 The proton non- flip and helicity-flip amplitudes for
longitudinally polarized photons ${\cal M}^{M}_{0\nu',0\nu}$,
which dominates at large $Q^2$, can be written in the form:
\begin{eqnarray}\label{pip}
 {\cal M}^{M}_{0+,0+} &\propto& \sqrt{1-\xi^2}\,
                             \,[- \frac{\xi
  (m_{N^i}+M_{N^f}) Q^2}{1-\xi^2}\frac{\rho_M}{t-m_M^2}+\langle \tilde{H}^{M}\rangle
  - \frac{\xi^2}{1-\xi^2} \langle
\widetilde{E}^{M}_{n.p.}\rangle];\;\nonumber\\
 {\cal M}^{M}_{0-,0+} &\propto&
\frac{\sqrt{-t^\prime}}{(m_{N^i}+M_{N^f})}\,\Big[
(m_{N^i}+M_{N^f}) Q^2\frac{\rho_M}{t-m_M^2}+ \xi \langle
\widetilde{E}^{M}_{n.p.}\rangle \Big]\,.
\end{eqnarray}
 Here $M$- produced
pseudoscalar meson, $N^i$-initial nucleon (proton), $N^f$-final
barion (neutron, $\Lambda$, $\Sigma$). The corresponding
amplitudes with transversally polarized photons are suppressed as
$1/Q$.

The first terms in (\ref{pip}) appear for the charged meson
production and are connected with the M meson pole. The fully
experimentally measured electromagnetic form factor of M meson is
included into $\rho_M$.

The second terms in (\ref{pip}) represent the handbag contribution
to the PML amplitude. The $<\tilde{F}>$ in (\ref{pip}) is a
convolution of GPD $\tilde F$ with the hard subprocess amplitude
${\cal H}_{0\lambda,0\lambda}(\xb,...)$:
\begin{equation}\label{ff}
<\tilde{F}>= \sum_\lambda\int_{-1}^1 d\xb
   {\cal H}_{0\lambda,0\lambda}(\xb,...) \tilde{F}(\xb,\xi,t),\;
\end{equation}

The subprocess amplitude is calculated within the MPA
\cite{sterman}. The  amplitude ${\cal H}^a$ is   a contraction of
the hard part ${\cal F}^a$, which is calculated perturbatively and
includes the transverse quark momentum $\vk$, and the
nonperturbative $\vk$-dependent meson wave function $\Psi$
\cite{koerner}. The gluonic corrections are treated in the form of
the Sudakov factors. The resummation and exponentiation of the
Sudakov corrections $S$ can be done  in the impact parameter space
$\vbs$ \cite{sterman}.  The Fourier transformed  subprocess
amplitude from the $\vk$ to $\vbs$ space  reads as
\begin{eqnarray}\label{aa}
{\cal H}^a_{0\lambda,0\lambda}\propto \int d\tau d^2b\,
         {\Psi}(\tau,-\vbs)\,\nonumber
         {\cal F}^{a}_{0\lambda,0\lambda}(\xb,\xi,\tau,Q^2,\vbs,)\,
      \als\, {\rm
      exp}{[-S(\tau,\vbs,Q^2)]}.
\end{eqnarray}
Here $\tau$ is the momentum fraction of the quark  that enters
into the meson.

The GPDs are estimated using the double distribution
representation \cite{mus99} which connects  GPDs with PDFs through
the double distribution function $f$. For the valence quark
contribution it looks like
\begin{eqnarray}\label{ddf}
f_i(\beta,\alpha,t)= h_i(\beta,t)\,
                   \frac{3}{4}\,
                   \frac{[(1-|\beta|)^2-\alpha^2]}
                           {(1-|\beta|)^{3}}.
\end{eqnarray}
The functions $h$  are determined in the terms of PDFs and are
parameterized in the form
\begin{equation}\label{pdfpar}
h(\beta,t)= N\,e^{b_0 t}\beta^{-\alpha(t)}\,(1-\beta)^{n}.
\end{equation}
Here  the $t$- dependence is considered in a Regge form and
$\alpha(t)$ is the corresponding Regge trajectory. The parameters
in (\ref{pdfpar}) are obtained from the known information about
PDFs \cite{CTEQ6} e.g, or from the nucleon form factor analysis
\cite{pauli}.

We calculate the leading-twist amplitudes  together with the meson
pole contribution on the basis of  (\ref{pip}). Unfortunately,
these terms are insufficient to describe experimental data at low
$Q^2$. We can demonstrate this using the $A_{UT}$ asymmetry in the
$\pi^+$ leptoproduction as an example.

\begin{figure}[h!]
\begin{center}
\mbox{\epsfig{figure=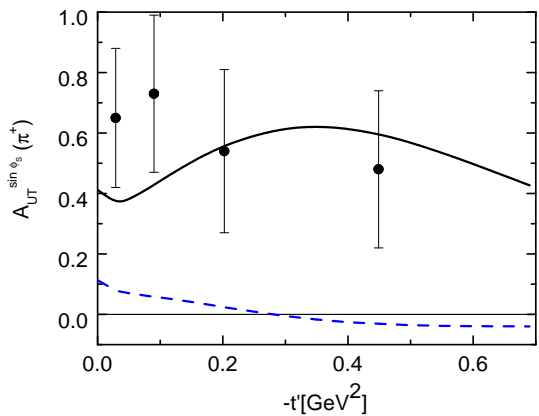, width=7cm}}
\end{center}
\caption{$A_{UT}^{\sin(\phi_s)}$ asymmetry of the $\pi^+$
production. Dashed line- leading twist contribution. Solid line-
model results, including twist-3 effects. Data are from HERMES
\cite{airap}}
\end{figure}

This asymmetry is expressed in terms of interference of
$M_{0-,++}$ and proton non-flip amplitude \cite{gk09}:
\begin{equation}\label{aut}
A_{UT}^{\sin(\phi_s)} \propto \mbox{Im}[ M^*_{0-,++} M_{0+,0+}].
\end{equation}

The leading twist contributions cannot explain this asymmetry
--see Fig. 2. A new twist-3 contribution to the $M_{0-,++}$
amplitude, which is not small at $t' \sim 0$, is needed. We
estimate  this contribution to ${\cal M}_{0-,++}$ by the
transversity GPD $H_T$ in conjugation with the twist-3 pion wave
function in the hard subprocess amplitude ${\cal H}$ \cite{gk11}.
We have
\begin{equation}\label{ht}
{\cal M}^{M,twist-3}_{0-,\mu+} \propto \,
                            \int_{-1}^1 d\xb
   {\cal H}_{0-,\mu+}(\xb,...)\,[H^{M}_T+...O(\xi^2\,E^M_T)].
\end{equation}
The $H_T$ GPD is connected with transversity PDFs  as
\begin{equation}
  H^a_T(x,0,0)= \delta^a(x);\;\;\; \mbox{and}\;\;\;
\delta^a(x)=C\,N^a_T\, x^{1/2}\, (1-x)\,[q_a(x)+\Delta q_a(x)].
\end{equation}
Here $a$ is a quark flavor. We parameterize the PDF $\delta$ by
using the model \cite{ans}. The double distribution (\ref{ddf}) is
used to calculate GPD $H_T$.

The amplitude $M_{0+,++}$ is extremely important in analyzes of
PML as well. The  transversity twist-3 contribution to this
amplitude is determined by $\bar E_T$ GPDs and has the form
\cite{gk11} similar to (\ref{ht})
\begin{equation}\label{et}
{\cal M}^{M,twist-3}_{0+,\mu+} \propto \, \frac{\sqrt{-t'}}{4 m}\,
                            \int_{-1}^1 d\xb
 {\cal H}_{0-,\mu+}(\xb,...)\; \bar E^{M}_T.
\end{equation}
The hard scattering subprocess amplitude ${\cal
H}_{0-,\mu+}(\xb,...)$ in (\ref{et}) is the same as in (\ref{ht}).

\begin{figure}[h!]
\begin{center}
\begin{tabular}{cc}
\includegraphics[width=6.1cm,height=5cm]{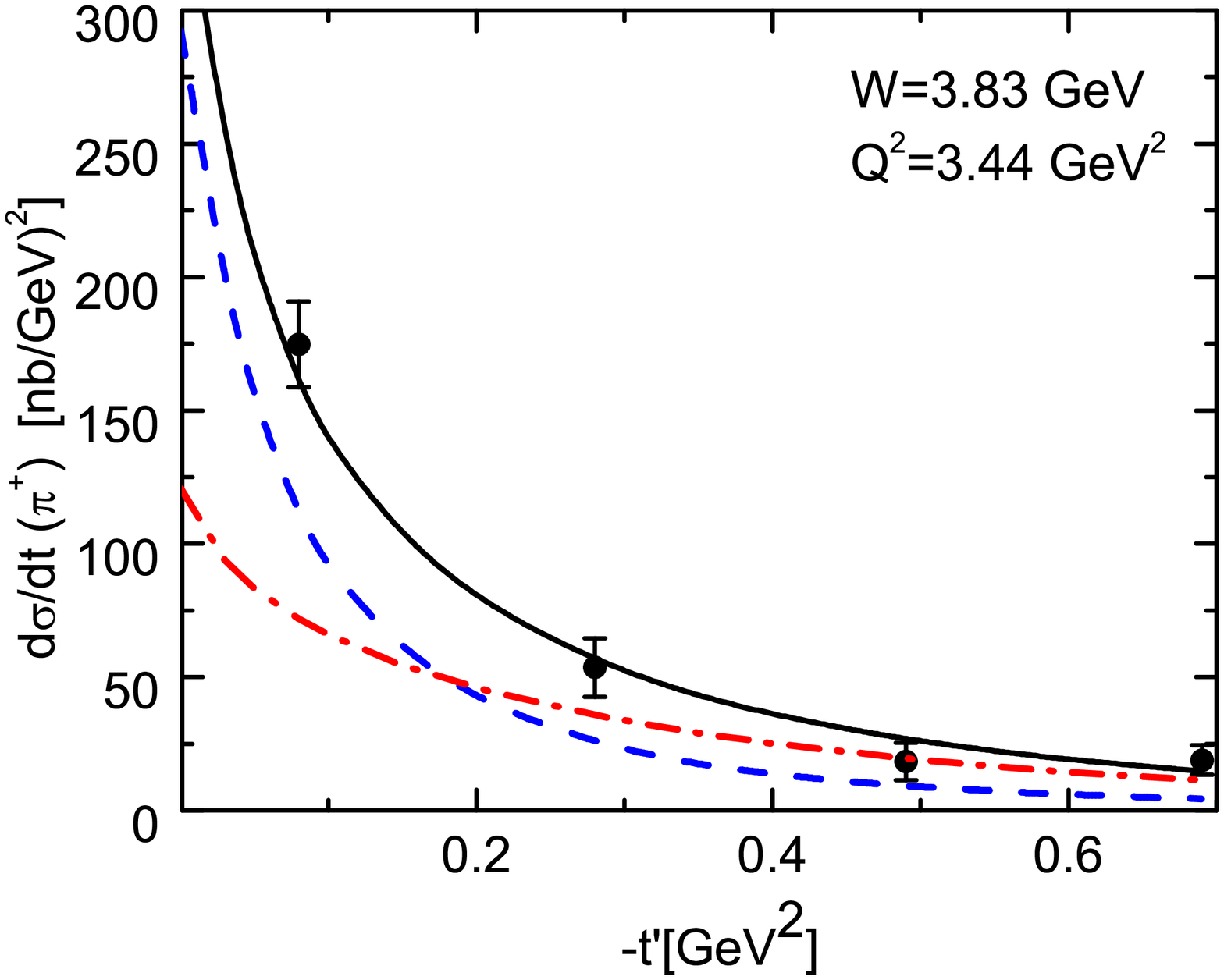}&
\includegraphics[width=6.1cm,height=5cm]{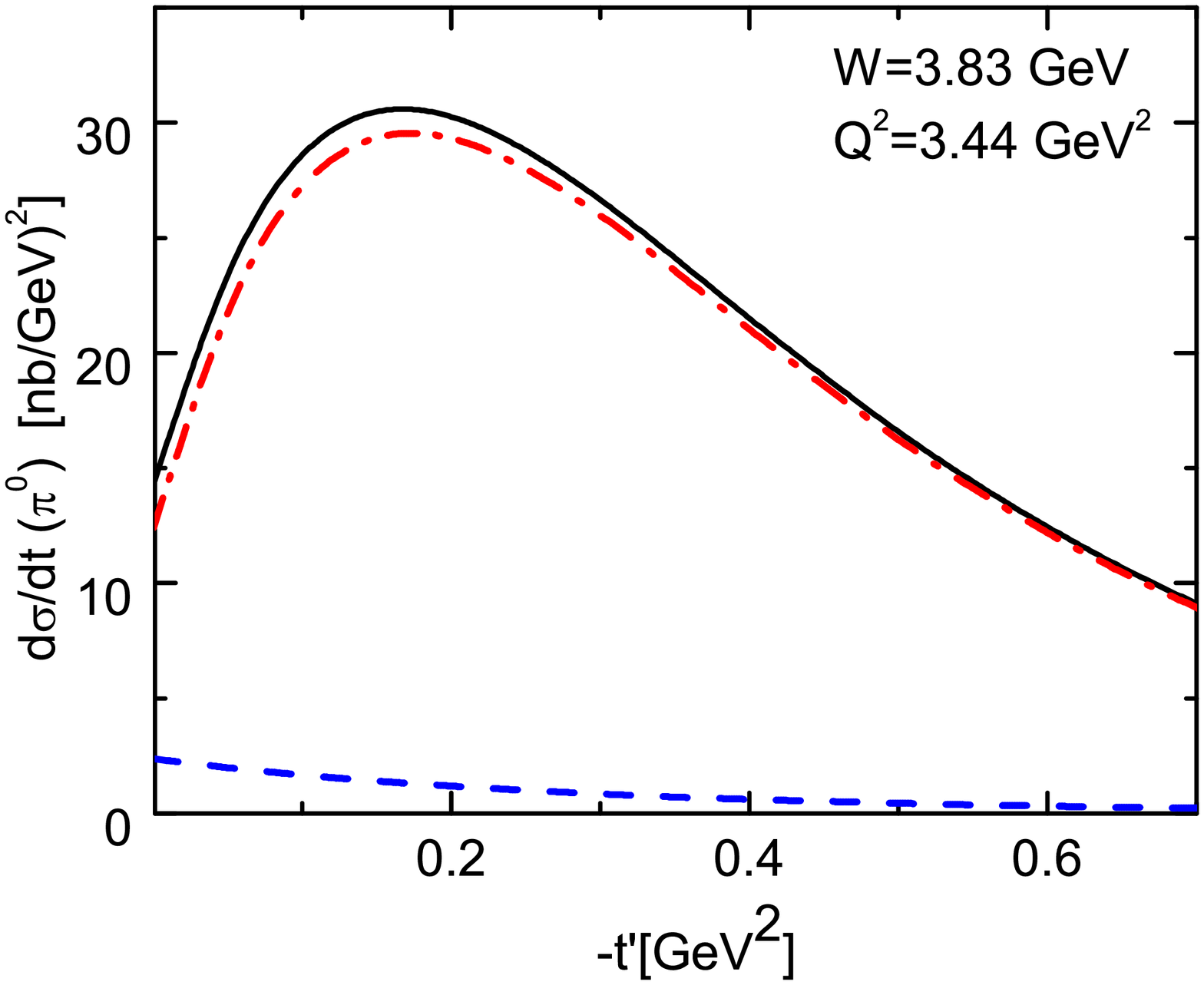}
\end{tabular}
\end{center}
\caption{Left: Cross section of $\pi^+$ production at HERMES.
Right: Cross section of $\pi^0$ production at HERMES. Full line-
unseparated cross section. dashed- $d \sigma_L/dt$, dashed-dotted
line- $d \sigma_T/dt$.}
\end{figure}

At the moment, the information on $\bar E_T$ is very poor. Some
results were obtained only in the lattice QCD \cite{lat}. The
lower moments of $\bar E_T^u$ and $\bar E_T^d$ were found to be
quite large, have the same sign and a similar size. At the same
time, $H_T^u$  and $H_T^d$ are different in  sign. For the pion
production we have the following contribution to GPDs
\cite{frankfurt99}
\begin{eqnarray}\label{gpi}
F(\pi^+)&=&F^{(3)}=F^u-F^d,\nonumber\\
F(\pi^0)&=&2/3\, F^u+ 1/3\,F^d.
\end{eqnarray}
From these equations  we find an essential compensation of the
$\bar E_T$ contribution to the $\pi^+$ amplitude but $H_T$ effects
are not small there. For the $\pi^0$ production we have the
opposite case -- $\bar E_T$ contributions are large but $H_T$
effects are smaller.

In Fig. 3 (left), we show our results \cite{gk09} for the
unseparated cross section of the $\pi^+$ production which
describes fine HERMES data \cite{airapsig}. The $\sigma_L$ and
$\sigma_T$ are shown as well. The longitudinal cross section
determined by leading-twist contribution dominates at small
momentum transfer $-t < 0.2 \mbox{GeV}^2$. At larger $-t$ we find
a not small transverse cross section where the $H_T$ contribution
is visible. In Fig. 3 (right), our results for the cross section
of the $\pi_0$ production are presented which are very different
from the $\pi^+$ process. The transverse cross section
\begin{equation}\label{st}
\sigma_T \propto |{\cal M}^{M,twist-3}_{0+,++}|^2+ |{\cal
M}^{M,twist-3}_{0-,++}|^2,
\end{equation}
where the $\bar E_T$ and $H_T$ contributions are important
\cite{gk09} dominates. At small momentum transfer the $H_T$
contribution is visible and provides a nonzero cross section. At
larger $-t' \sim 0.2 \mbox{GeV}^2$ the $E_T$ contribution is
essential and gives a maximum in the cross section.

The longitudinal cross section which is expected to play an
important role  is much smaller with respect to the transverse
cross section $\sigma_T$. The essential contributions to the
$\sigma_T$ cross section are determined by the twist-3 $H_T$ and
$\bar E_T$ effects and decreases quickly with $Q^2$. At quite
large $Q^2$ the leading-twist effects will dominate.

\begin{figure}[h!]
\begin{center}
\begin{tabular}{cc}
\includegraphics[width=6.1cm,height=5cm]{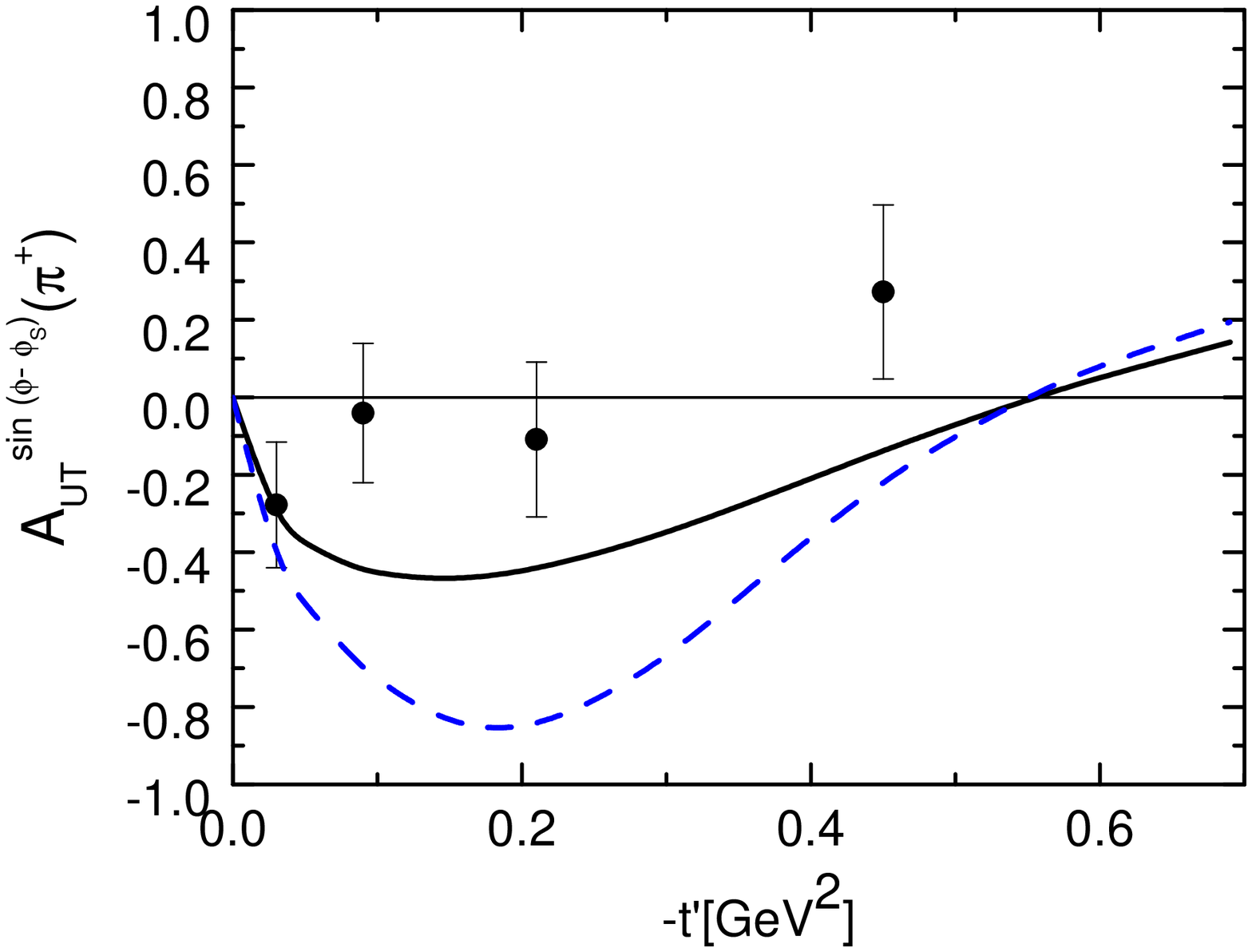}&
\includegraphics[width=6.1cm,height=5cm]{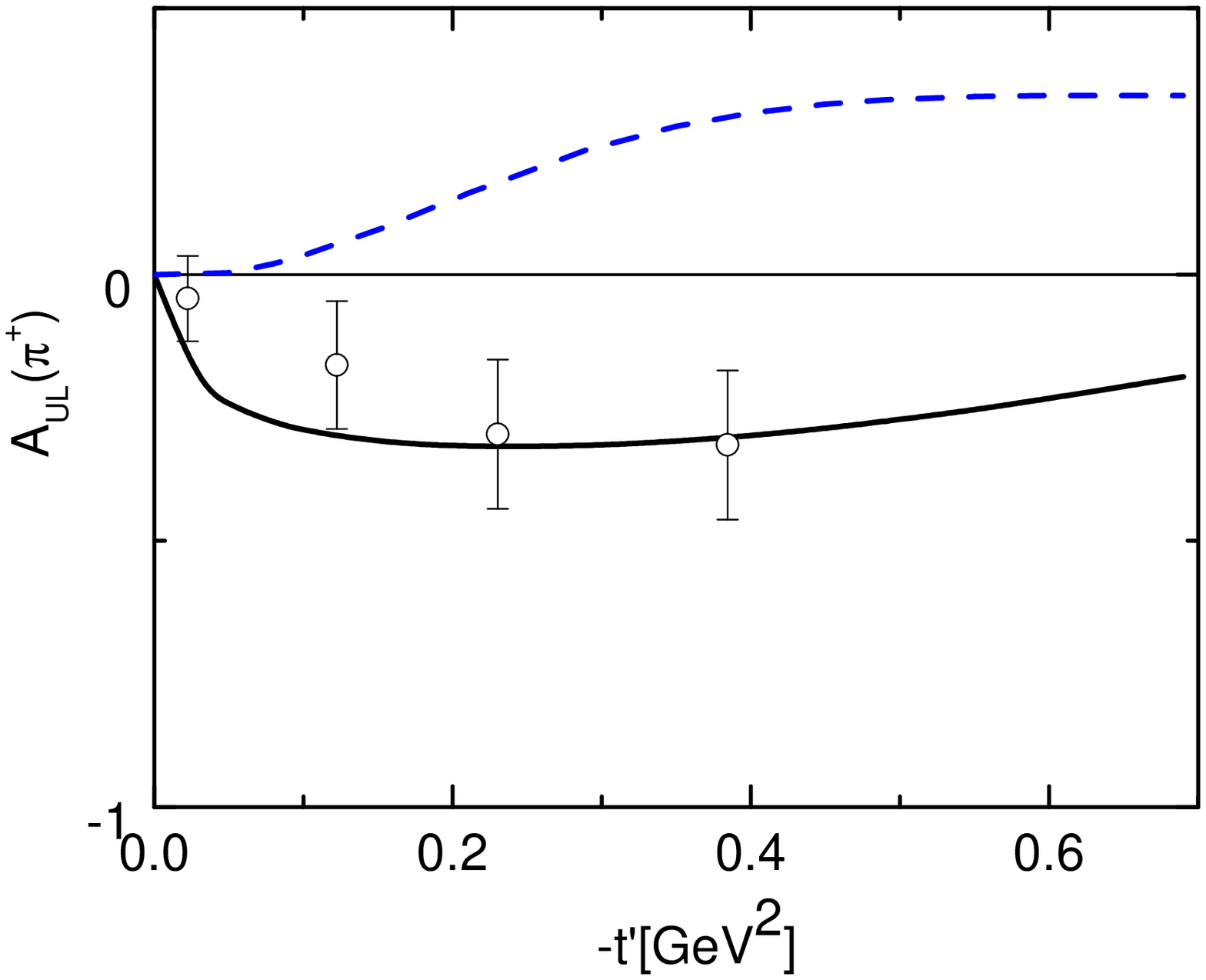}
\end{tabular}
\end{center}
\caption{Left: $A_{UT}$ asymmetry of the $\pi^+$ production at
HERMES energies. Right: $A_{UL}$ asymmetry of the $\pi^+$
production at HERMES. HERMES data are shown \cite{airap}. Dashed
line- results without transversity $H_T$ effects.}
\end{figure}

In Fig.4, we demonstrate that the transversity $H_T$ effects are
essential in asymmetries of the $\pi^+$ production. When we omit
the $H_T$ contributions, asymmetries change drastically.

In Fig 5 (left), our prediction for the $\pi^0$ production in the
CLAS energy range \cite{sgspin11} is shown together with
experimental data \cite{bedl}. Our results are close  to the
experimental data and definitely show the same dip in the
unseparated cross section at low momentum transfer, as was
observed for HERMES --see Fig.3 (right). We present in this plot
the interference $\sigma_{LT}$ and $\sigma_{TT}$ cross sections
too. The value of $\sigma_{LT}$ is quite small, compatible with
zero. The $\sigma_{TT}$ cross section is negative and large. Note
that the $E_T$ contribution to $\sigma_{T}$ and $\sigma_{TT}$
cross sections is strongly correlated. The fact that we describe
the CLAS data for both cross sections quite well can be an
indication of observation of large transversity effects at CLAS.
However, the definite conclusion on the importance of transversity
effects in the $\pi^0$ cross section can be made only if the data
on the separated $\sigma_L$ and $\sigma_T$ cross section will be
available experimentally and $\sigma_{T}$ will be much larger than
$\sigma_{L}$. Probably, such a study can be performed at JLAB12.

\begin{figure}[h!]
\begin{center}
\begin{tabular}{cc}
\includegraphics[width=6.1cm,height=5cm]{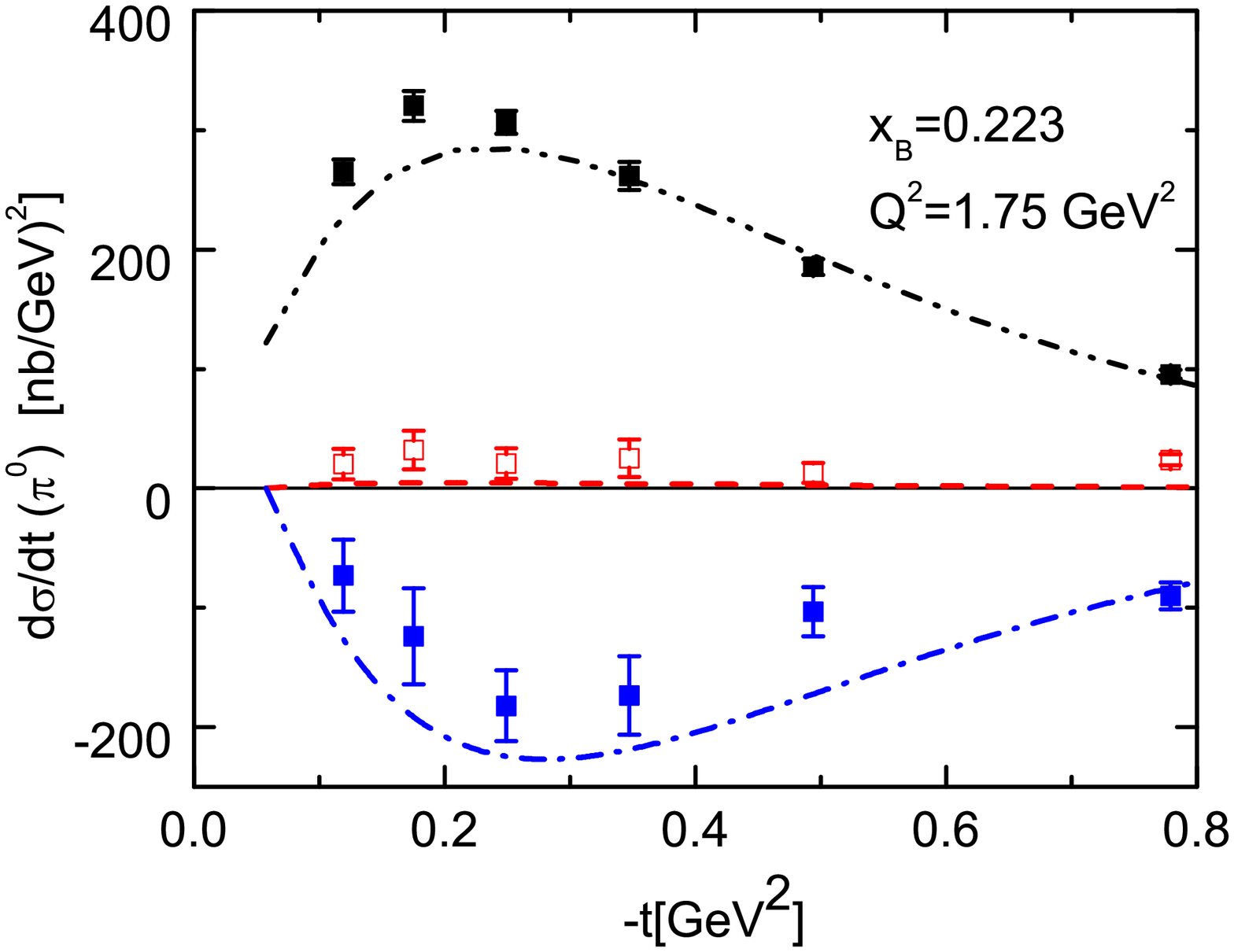}&
\includegraphics[width=6.1cm,height=5cm]{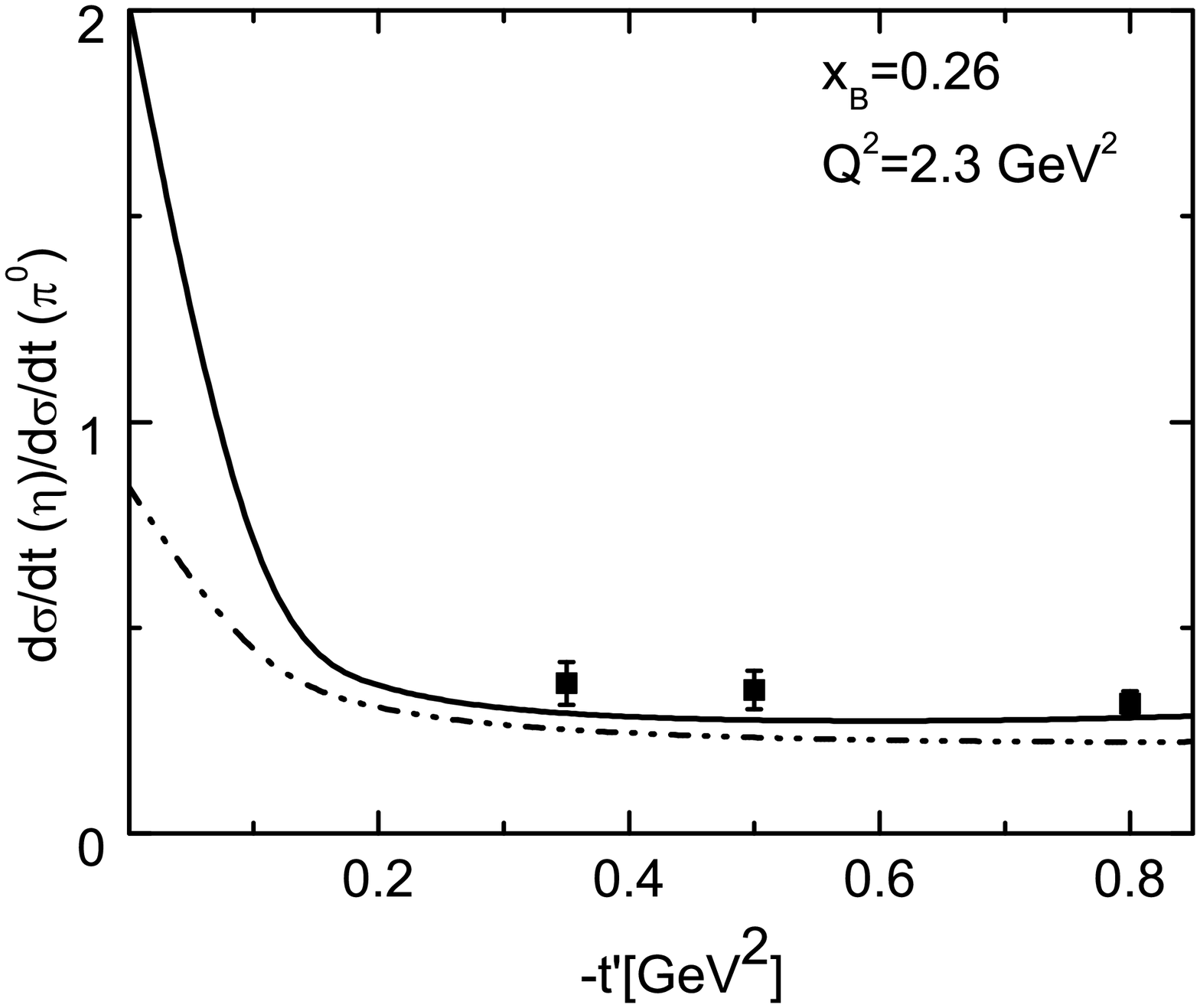}
\end{tabular}
\end{center}
\caption{Left: $\pi^0$ production in the CLAS energy range
together with the data. Dashed-dot-dotted line-
$\sigma_{T}+\epsilon \sigma_{L}$, dashed line-$\sigma_{LT}$,
dashed-dotted- $\sigma_{TT}$. Right: $\eta/\pi^0$ production ratio
in the CLAS energy range together with preliminary data.}
\end{figure}

In Fig. 5 (right), we analyze the transversity effects in the
ratio of the $\eta/\pi^0$ cross section at CLAS energies.  The two
parameterizations of $H_T$ GPDs \cite{gk11} are presented  there.
Different combinations of the quark contributions to these
processes lead to the essential role of $H_T$ effects in this
ratio at small $-t <0.2 \mbox{GeV}^2$. At larger momentum transfer
large $E_T$ effects in the $\pi^0$ production found in the model
lead to a rapid decrease of the $\eta/\pi^0$ cross section ratio
with $t$- growing. At $-t
>0.2 \mbox{GeV}^2$ this ratio becomes close to $\sim 0.3$, which
was confirmed by CLAS \cite{vkubar}.

\begin{figure}[h!]
\begin{center}
\begin{tabular}{cc}
\includegraphics[width=6.1cm,height=5cm]{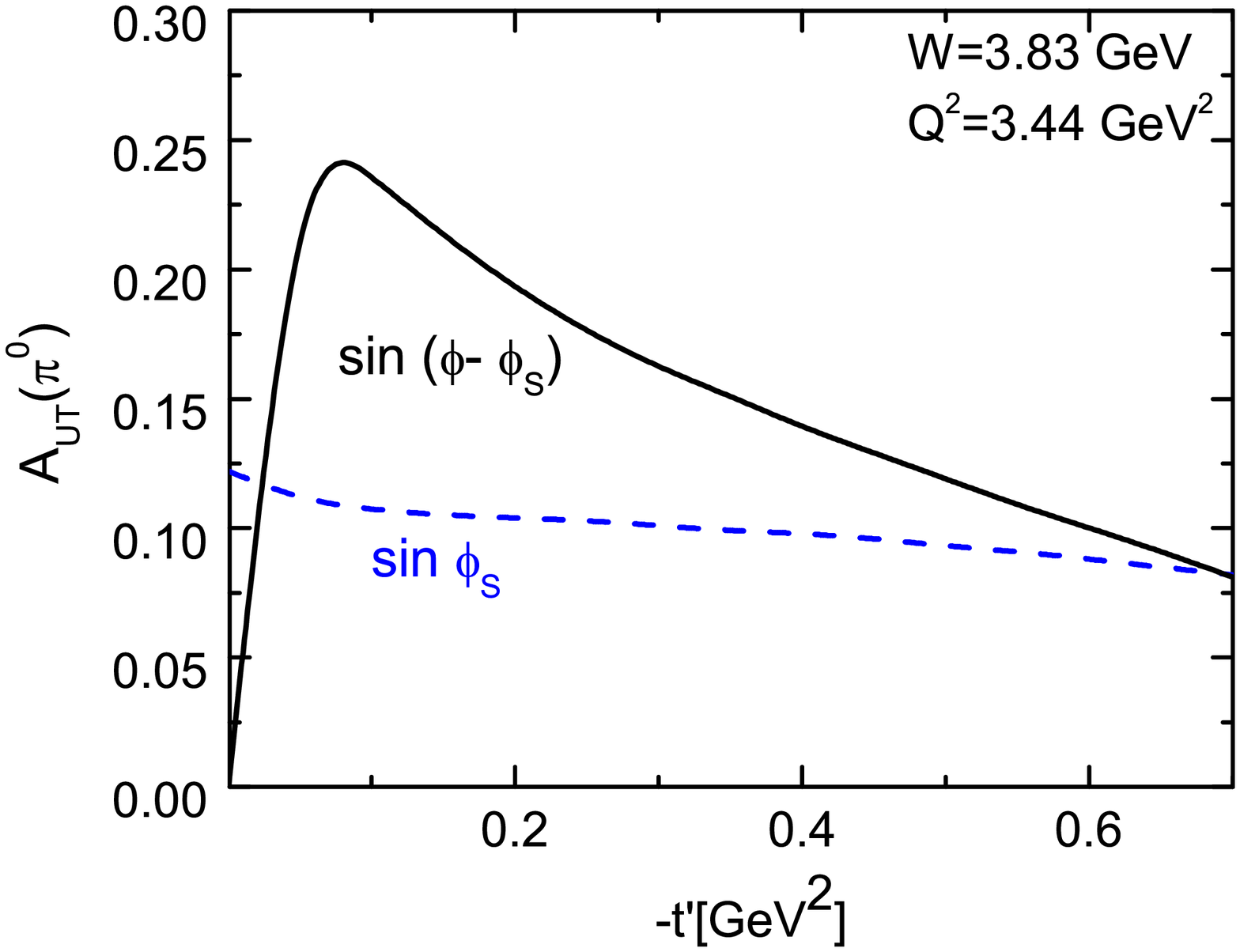}&
\includegraphics[width=6.1cm,height=5cm]{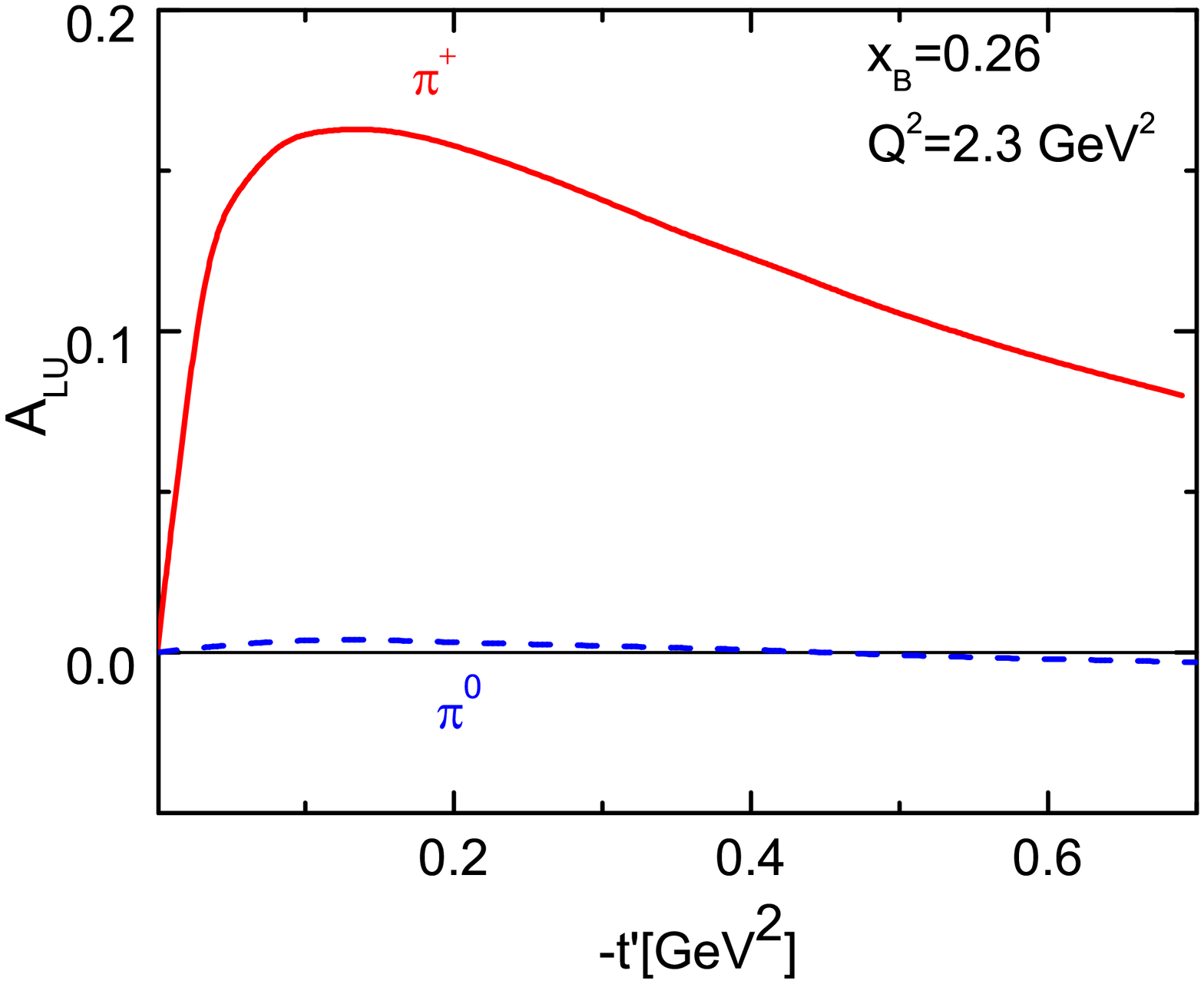}
\end{tabular}
\label{fig:4}
\end{center}
\caption{Left: Moments of the $A_{UT}$ asymmetries at HERMES for
the $\pi^0$ production.  Right: The predicted $A_{LU}$ asymmetry
in the $\pi^+$ and $\pi^0$  production at HERMES.}
\end{figure}

In Fig.6 (left), we present our results for the moments of the
$A_{UT}$ asymmetry in the $\pi^0$ production at HERMES. The
predicted asymmetries are large and can give  additional
information on transversity effects in this reaction. In Fig.6
(right), we show the $A_{LU}$ asymmetry in the pion production at
HERMES. $A_{LU}(\pi^+)$ is large because of the pion-pole
contribution in this channel. The predicted $A_{LU}$ asymmetry in
the $\pi^0$ production is small. Measurement of this asymmetry at
HERMES can give information on the nonpole term of
$\widetilde{E}^{M}_{n.p.}$ in (\ref{pip}).

\begin{figure}[h!]
\begin{center}
\begin{tabular}{cc}
\includegraphics[width=6.1cm,height=5cm]{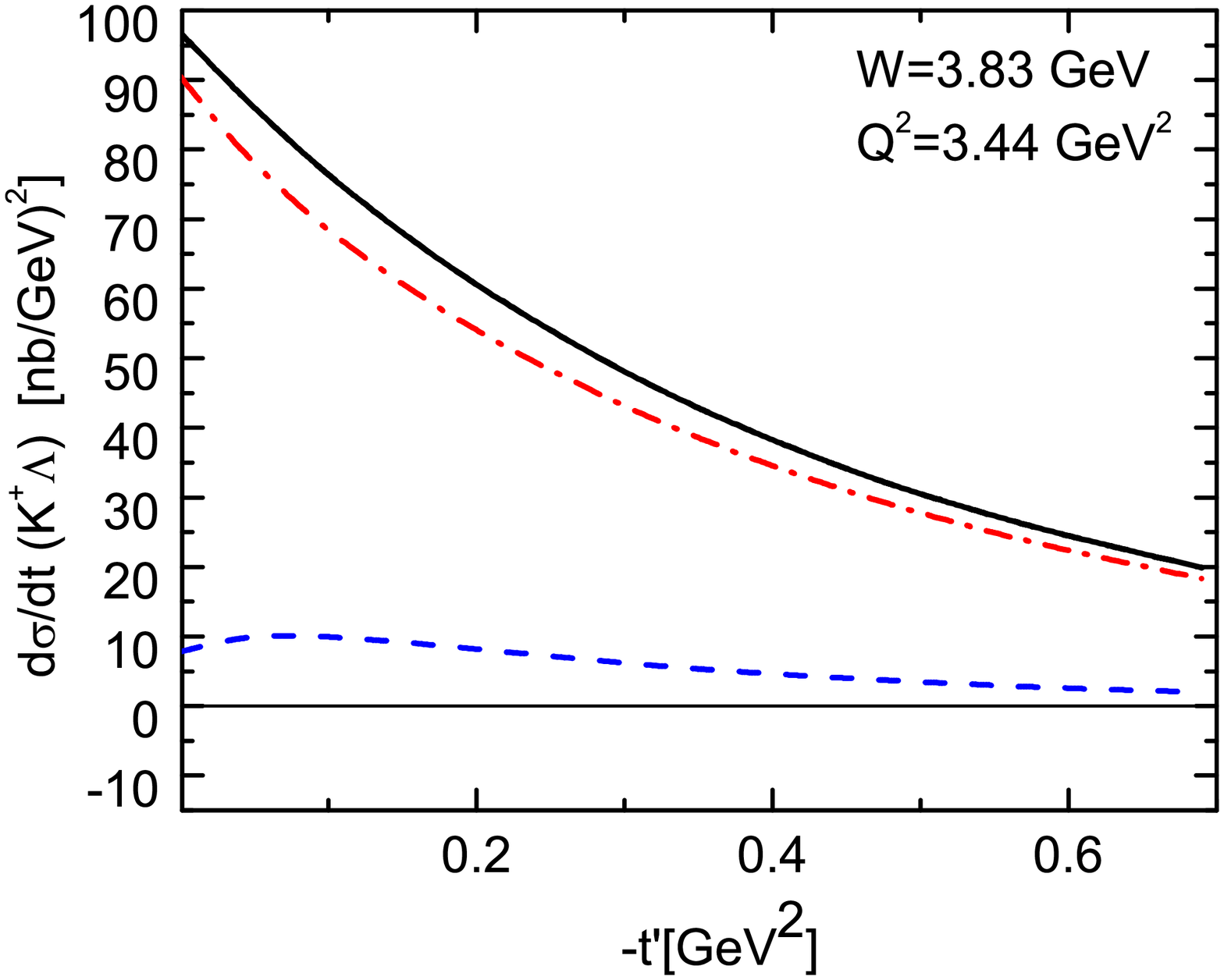}&
\includegraphics[width=6.1cm,height=5cm]{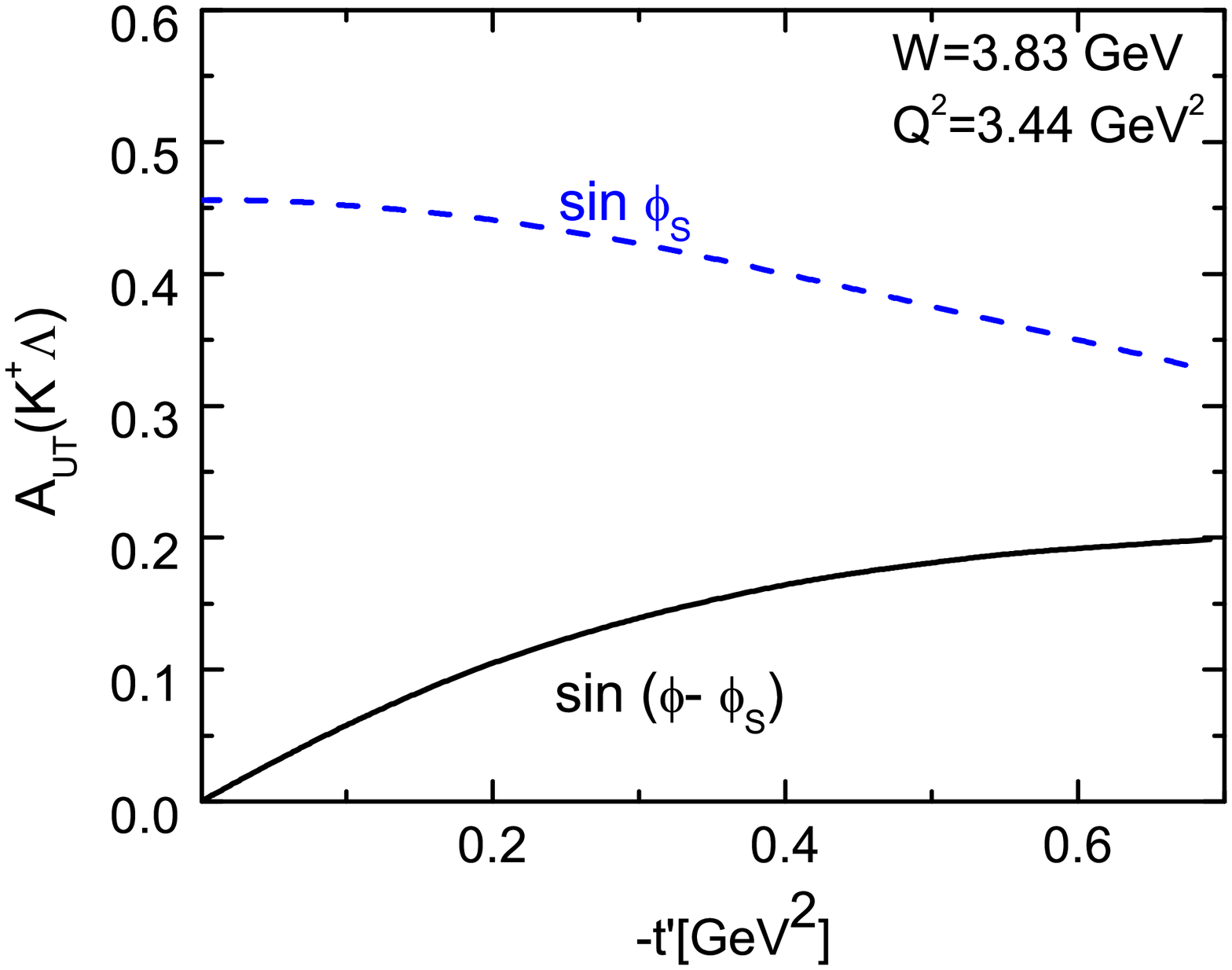}
\end{tabular}
\label{fig:5}
\end{center}
\caption{Left: The $K^+ \Lambda$ production cross sections at
HERMES energies. Right: Predicted  moments of $A_{UT}$ asymmetries
for $K^+ \Lambda$ channel at HERMES.}
\end{figure}

Using the same model we  calculate the cross section and spin
asymmetry for the $K^+ \Lambda$ production. To estimate proton-
hyperon transition GPDs we use  the SU(3) flavor symmetry model
\cite{frankfurt99}
\begin{equation}
  H_T(p \to \Lambda) \sim [2
H_T^u-H_T^d-H_T^s].
\end{equation}
Due to  different signs of $H_T^u$ and $H_T^d$ we find  a quite
large $H_T$ effect here. In this reaction, the kaon pole
contribution should be much smaller with respect to the $\pi^+$
case. The details of calculations can be found in \cite{gk11}. The
large transversity $H_T$ effects in the $K^+ \Lambda$ channel
provide the large $\sigma_T$ cross section without a forward dip
which dominated with respect to $\sigma_L$, see Fig. 7 (left). In
Fig. 7 (right), we show our predictions for moments of the
$A_{UT}$ asymmetry in this channel. The $\sin(\phi_s)$ moment of
asymmetry determined by the $H_T$ transversity contribution
(\ref{aut}) is quite large.

\section{Conclusion}
We  calculate the PML amplitude  within the handbag approach, in
which the amplitudes factorize into hard subprocesses and  GPDs
\cite{fact}. The hard subprocess amplitudes were calculated within
the modified perturbative approach \cite{sterman} where quark
transverse degrees of freedom and the gluonic radiation, condensed
in a Sudakov factor were taken into account.

 At leading-twist accuracy the PML reactions are
sensitive to the GPDs $\widetilde{H}$ and $\widetilde{E}$ which
contribute to the amplitudes for longitudinally polarized virtual
photons. This contribution should be predominated at large $Q^2$.
Unfortunately, now experimental data on these reactions are
available at small photon virtualities.

 We observed that the experimental data on
pseudoscalar meson leptoproduction at low $Q^2$ also require
contributions from the transversity GPDs, in particular, from
$H_T$ and $\bar E_T$. Within the handbag approach the transversity
GPDs are accompanied by a twist-3 meson wave function. At HERMES
and COMPASS energies the twist-3 $\bar E_T$ effects produce a
large transverse cross section $\sigma_T$ \cite{gk11} which
exceeds substantially the leading twist longitudinal cross section
for most reactions with the exception of the $\pi^+$ and $\eta'$
channels.

The indication of large transversity effects are available now at
CLASS. They observe a large unseparated  and large negative
$\sigma_{TT}$ cross section which can be described in our model by
large transversity $\bar E_T$ effects. Essential $H_T$ and $\bar
E_T$ effects are predicted at the ratio $\eta/\pi^0$ cross
section. Large $E_T$ effects in the $\pi^0$ production predict
that for $-t>0.2 \mbox{GeV}^2$ this ratio should be close to $\sim
0.3$, which was confirmed by CLAS \cite{vkubar}.

Nevertheless the experimental separation of the $\sigma_L$ and
$\sigma_T$ cross section in the $\pi^0$ electroproduction is
important. If it is found that $\sigma_{T}$ is  much larger than
the $\sigma_{L}$ cross section, this will be a definite
demonstration of observation of transversity effects in this
reaction. We hope that it can be done at JLAB12. Essential $H_T$
effects in the $K^+ \Lambda$ channel were predicted.

We describe well the cross section and spin observables for
various PML. Thus, we can conclude that the information on GPDs
discussed above should be  not far from reality. Future
experimental results at COMPASS, JLAB12 can give  important
information on the role of transversity effects in these
reactions.
\bigskip

This work is supported  in part by the Russian Foundation for
Basic Research, Grant  12-02-00613  and by the Heisenberg-Landau
program.

\end{document}